\documentclass[nonacm,sigplan]{acmart}

\usepackage[]{hyperref}

\usepackage{titlesec}
\usepackage{comment}
\usepackage{amsmath,newtxmath}
\usepackage{array}
\usepackage{multirow}
\usepackage{graphicx}
\usepackage{subcaption}
\usepackage{color}
\usepackage{algorithm}
\usepackage{algpseudocode}
\usepackage{pifont}
\usepackage{threeparttable}
\usepackage{makecell}
\usepackage{booktabs} 

\usepackage{tikz}
\usepackage{xcolor}
\newcommand*\circled[1]{\tikz[baseline=(char.base)]{
            \node[shape=circle,fill,inner sep=0.8pt] (char) {\textcolor{white}{#1}};}}

\usepackage{listings}

\definecolor{codegreen}{rgb}{0,0.6,0}
\definecolor{codegray}{rgb}{0.5,0.5,0.5}
\definecolor{codepurple}{rgb}{0.58,0,0.82}
\definecolor{backcolour}{rgb}{0.95,0.95,0.92}

\lstdefinestyle{mystyle}{
    backgroundcolor=\color{backcolour},   
    commentstyle=\color{codegreen},
    keywordstyle=\color{magenta},
    numberstyle=\tiny\color{codegray},
    stringstyle=\color{codepurple},
    basicstyle=\ttfamily\footnotesize,
    breakatwhitespace=false,         
    breaklines=true,                 
    captionpos=b,                    
    keepspaces=true,                 
    numbers=left,                    
    numbersep=5pt,                  
    showspaces=false,                
    showstringspaces=false,
    showtabs=false,                  
    tabsize=2
}

\newcommand{\comments}[2]{{\bf[\textcolor{red}{#1}: \textcolor{blue}{#2}]}}

\newcommand{\CZ}[1]{{\comments{Chen}{#1}}}

\newcommand{\para}[1]{{\bf \noindent #1 \hspace{2pt}}}
\newcommand{\todo}[1]{{\color{red}[\textbf{\sc TODO}: \textbf{#1}]}}
\newcommand{\hl}[1]{{\color{red}\textbf{#1}}}
\newcommand{\oursys}{{{Infinite-LLM}}}

\begin{document}

\newcommand{\attename}{{DistAttention}}
\newcommand{\sysname}{{Infinite-LLM}}
\title{\oursys{}: Efficient LLM Service for Long Context with DistAttention and Distributed KVCache}

\author{Bin Lin}
\authornote{Equal contribution.}
\affiliation{%
  \institution{Alibaba Group}
  \country{}
}

\author{Chen Zhang}
\authornotemark[1]
\authornote{Corresponding author: chenzhang.sjtu@sjtu.edu.cn.}
\affiliation{%
  \institution{Shanghai Jiao Tong University}
  \country{}
}

\author{Tao Peng}
\authornotemark[1]
\affiliation{%
  \institution{Alibaba Group}
  \country{}
}

\author{Hanyu Zhao}
\affiliation{%
  \institution{Alibaba Group}
  \country{}
}

\author{Wencong Xiao}
\affiliation{%
  \institution{Alibaba Group}
  \country{}
}

\author{Minmin Sun}
\affiliation{%
  \institution{Alibaba Group}
  \country{}
}

\author{Anmin Liu}
\affiliation{%
  \institution{Peking University}
  \country{}
}

\author{Zhipeng Zhang}
\affiliation{%
  \institution{Alibaba Group}
  \country{}
}

\author{Lanbo Li}
\affiliation{%
  \institution{Alibaba Group}
  \country{}
}

\author{Xiafei Qiu}
\affiliation{%
  \institution{Alibaba Group}
  \country{}
}

\author{Shen Li}
\affiliation{%
  \institution{Alibaba Group}
  \country{}
}

\author{Zhigang Ji}
\affiliation{%
  \institution{Shanghai Jiao Tong University}
  \country{}
}

\author{Tao Xie}
\affiliation{%
  \institution{Peking University}
  \country{}
}

\author{Yong Li}
\affiliation{%
  \institution{Alibaba Group}
  \country{}
}

\author{Wei Lin}
\affiliation{%
  \institution{Alibaba Group}
  \country{}
}

\begin{abstract}


Large Language Models (LLMs) demonstrate substantial potential across a diverse array of domains via request serving. However, as trends continue to push for expanding context sizes, the autoregressive nature of LLMs results in highly dynamic behavior of the attention layers, showcasing significant differences in computational characteristics and memory requirements from the non-attention layers. This presents substantial challenges for resource management and performance optimization in service systems. Existing static model parallelism and resource allocation strategies fall short when dealing with this dynamicity.
To address the issue, we propose \oursys{}, a novel LLM serving system designed to effectively handle dynamic context lengths. \oursys{} disaggregates attention layers from an LLM's inference process, facilitating flexible and independent resource scheduling that optimizes computational performance and enhances memory utilization jointly. By leveraging a pooled GPU memory strategy across a cluster, \oursys{} not only significantly boosts system throughput but also supports extensive context lengths. 
Evaluated on a dataset with context lengths ranging from a few to 2000K tokens across a cluster with 32 A100 GPUs, \oursys{} demonstrates throughput improvement of 1.35-3.4x compared to state-of-the-art methods, enabling efficient and elastic LLM deployment.

\end{abstract}

\maketitle

\pagestyle{plain}

\section{Introduction}


Large Language Models (LLMs)\cite{chang2023survey, gpt35, palm2023,workshop2022bloom,llama3modelcard} have significantly advanced the field of generative artificial intelligence, and these inspiring capabilities have been integrated into various aspects of daily life. The universality of LLMs is evident across numerous domains, such as programming copilot\cite{chen2021evaluating, huang2023agentcoder}, document summarization\cite{zhang2024benchmarking, wang2023element}, information retrieval\cite{zhu2023large, tang2024self}, and chatbots\cite{wei2023leveraging, kim2023chatgpt}. The inference serving of LLMs~\cite{aminabadi2022deepspeed,li2023alpaserve,fastgen2023} has emerged as a critical component within cloud infrastructures.

Today’s LLM serving typically employs autoregressive mechanisms\cite{ bert, radford2018improving, brown2020language, sutskever2011generating} to iteratively generate output tokens and intermediate contexts (a.k.a.  KVCache\cite{pope2023efficiently}). The autoregressive nature of these models introduces a characteristic of unpredictability in the sequence of generated tokens, as the process continues until the generation of an end token. As a result, the required memory and computational resources for LLM services dynamically change, with neither the lifetime nor the length of the context known a priori. With the rapid development of LLMs, the supported context is continuously expanding\cite{anthropic}. Multiple LLM vendors have significantly increased their capacity to millions of tokens—e.g., 128K for ChatGPT \cite{openaichatgpt}, 1000K for Google's Gemini \cite{gemini}, and 2000K for LongRoPE \cite{ding2024longrope}.


In LLM cloud service systems, resource demands are highly dynamic due to the enormous dynamicity and unpredictability of context generation tasks for LLMs. Since context generation tasks may generate arbitrary lengths, from as few as 1 to up to 2000K tokens, the cloud services must cater to a broad range of demands. Due to the unpredictable length of context generated by each request, pre-assigning resources accurately becomes unfeasible, leading to highly varied demands for computing and memory resources. For example, a single instance (i.e., a model replica deployed to handle request
data in parallel)  might manage numerous compute-intensive short-context tasks at one time and switch to memory-intensive long-context tasks or a mix of varying lengths at another. 


The preceding complexity in dynamic resource demands results in LLM service systems that struggle to \textit{efficiently} and \textit{elastically} adapt to varying workload requirements under different context lengths. This often leads to reduced system efficiency, manifesting primarily in two aspects:

\textbf{Inefficient Model Parallelism inside an Instance.}
The model parallelism strategy required for processing requests with short and normal-length contexts differs significantly from that for long contexts. Traditional LLM service systems use a fixed model parallelism, where each instance is allocated a fixed number of GPUs. This fixed allocation makes it challenging to flexibly support both long and normal-length contexts efficiently. For example, processing a context with a length of 1K tokens on the Llama-7B model requires approximately 15 GB of memory, a fraction of an A100 GPU's capacity, while a context with a length of 1000K tokens demands over 500 GB, equivalent to the combined memory of about 7 A100 GPUs. Consequently, a higher degree of parallelism (DoP) is necessary for longer requests to meet their extensive resource needs, in stark contrast to the minimal requirements of shorter tasks. Configuring the system to meet the high DoP needed for long requests results in excessive model slicing and communication overhead for shorter requests, severely impacting performance. Existing parallelism strategies such as Tensor Parallelism and Pipeline Parallelism~\cite{megatronlm,pipeline}, which are based on static model dimensions, struggle to adapt flexibly to such dynamic workloads.

\textbf{Inefficient Resource Management across Instances.} The dynamic lengths of requests also limit the efficiency of resource management across instances and the cluster throughput. In particular, it is difficult for the scheduler to find an optimal request placement to saturate both memory and compute utilization. This is because the memory utilization is determined only by the total KVCache, whereas the compute utilization largely depends on the batch size, i.e., number of running requests. For example, when a request grows too long on an instance, its KVCache will consume too much memory space on that instance, which in turn will greatly reduce the running batch size and compute utilization, even the memory is saturated. Similarly, when requests are short on an instance, the spare memory also cannot be harvested by the long requests on other instances. As a result, the overall cluster throughput would be limited.

Through an in-depth analysis of the computational characteristics of LLM models, we identified that the root of the challenges lies in the significant differences between attention and non-attention layers: non-attention layers exhibit static behavior with changing sequence lengths and are sensitive to batch size, while attention layers display dynamic behavior and are not affected by the batch size. To address these challenges, we present \oursys{}, a novel LLM service system designed for managing highly dynamic context lengths in LLM requests. \oursys{} introduces a new approach that decouples the computation of attention layers from the rest of the LLM model. This decoupling allows for flexible and independent resource scheduling, specifically targeting the memory needs of dynamic attention layers and computation needs of the rest of the LLM model. Additionally, \oursys{} optimizes resource allocation by using the entire cluster's GPU memory as a pooled resource, allowing instances with surplus memory capacity to aid those processing extensive context tasks. This method not only significantly enhances resource efficiency and system throughput but also enables the cluster to support tasks with extremely long context lengths that surpass the memory limits of a single instance. 

The contributions of this paper are summarized as follows:

\begin{itemize}
    \item We reveal the dynamic characteristic of LLM request serving, and identify the limitations inherent in existing static model parallelism deployments and KVCache scheduling within a single instance.
    \item We present \attename{}, a novel attention mechanism that is mathematically equivalent to the original attention, designed to flexibly disaggregate attention computation and KVCache in a distributed way.
    \item We propose \oursys{}, an efficient LLM serving system specifically designed to adapt to the LLM serving dynamicity. It is capable of supporting scalable context length efficiently, by scheduling KVCache in cluster-level, thus to balance resource requirements between instances and achieve high overall system throughput. 
    \item Evaluations show that \oursys{} can serve 2,000K tokens with 32 GPUs, achieving 1.35-3.4× improvement in end-to-end performance compared to state-of-the-art LLM serving system.
\end{itemize}

\section{Background and Motivation}


\begin{figure*}[t!] 
    \centering
    \begin{minipage}{\textwidth}
        \centering
        \includegraphics[width=1.0\linewidth]{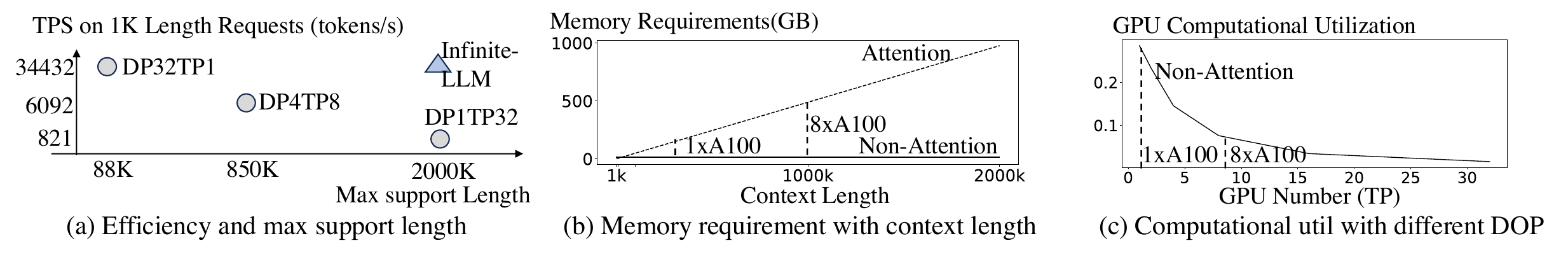}
        \caption{Static model parallelism struggles to maintain efficiency across all context length}
        \label{fig:challenge1}
    \end{minipage}
    \label{fig:test} 
\end{figure*}

\begin{figure*}[t!]
  \centering
  \includegraphics[width=0.97\linewidth]{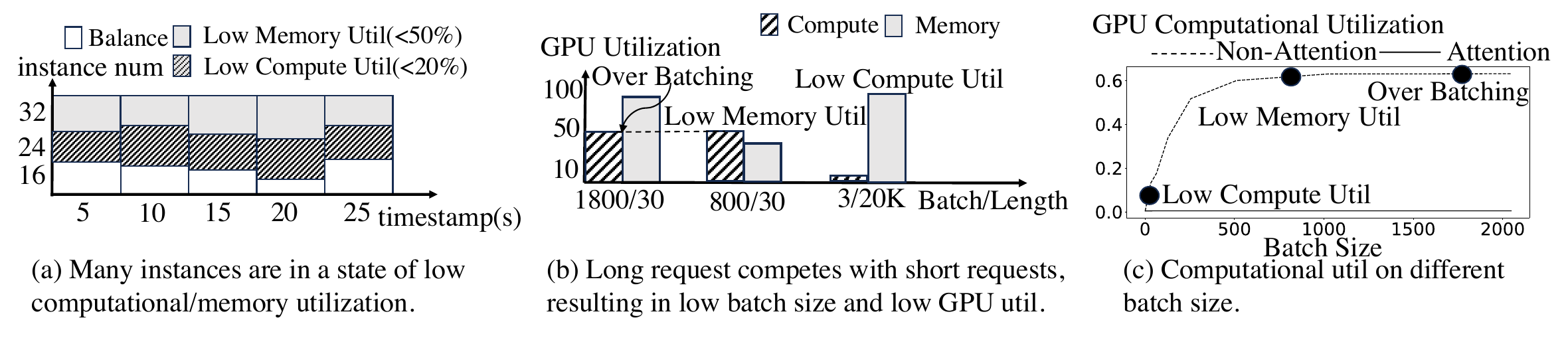}
  \caption{Resource under-utilization across instances in a cluster}
  \label{fig:challenge2}
\end{figure*}

\subsection{LLM Serving and Parallelism Method}
\para{LLM Inference.} Large Language Models (LLMs)\cite{chang2023survey, gpt35, palm2023} are dominated by Transformer architectures\cite{bert, brown2020language}.
A Transformer block consists of three key components: the {QKV Linear layer}, {Multi-Head Attention Mechanism}, and {Feed-Forward Neural Network (FFN)} modules. Multi-Head Attention involves attention kernels and persists key-value cache (KVCache) across iterations, while both the QKV Linear layers and FFN layers are mostly General Matrix to Matrix Multiplication (GEMM) kernels.
Inference serving of LLMs is autoregressive. Specially, the \textit{prefill} phase takes prompts as inputs to generate the first token, and each new token afterwards is generative iteratively until an ``end-of-sequence'' (EOS) token, usually referred to as \textit{decode}.
Due to the generality of LLM, context length of inference serving can be wide-ranging~\cite{adams2024longhealth,bai2023longbench,an2023eval,yuan2024lv}, from only 1  token~\cite{wang2024efficient} to 2000K tokens~\cite{ding2024longrope}.
To further scale the serving throughput capacity, multiple model replicas are deployed to handle request data in parallel (a.k.a. \textit{data-parallel}). Deployed as an \textit{instance}, each replica contains a full copy of model parameters. This instance is responsible for accommodating an LLM request and processing it to completion iteratively.

\para{Model Parallelism.}
Model-parallel is necessary for supporting large models that do not fit into the memory of a single device. It can expand the total memory available to the serving instance for storing its prompt inputs, model weights, and intermediate values. Tensor parallelism and pipeline parallelism are the two major categories.
Tensor-parallel~\cite{megatronlm} partitions a model layer across multiple GPUs, with each GPU executing a part of inference computation in parallel. Communication is required to split input or merge output for subsequent layers in the model. 
Pipeline-parallel\cite{narayanan2019pipedream, li2023alpaserve} avoids intra-layer communication by assigning contiguous layers on different GPUs for pipeline fashion execution, introducing inter-layer communication instead.

\subsection{System Design Challenges}

In this section, we delve into the computational characteristics of attention and non-attention layers at various text lengths and the challenges that these differences pose to system design. To better illustrate our points, we conduct a series of motivating experiments using the LLaMA2-7B model on 32 A100 GPUs.

\para{Observation 1: Instances with a higher number of GPUs are capable of supporting long-text tasks but perform poorly on normal-text tasks.} As illustrated in \autoref{fig:challenge1}(a), the performance differences across these instances are stark: Instance DP1TP32, with the most GPUs in a single instance (32 GPUs), has the largest memory capacity to support text generation tasks up to 2000K in length but performs the worst on  text generation tasks of standard length (1K). Conversely, as the number of GPUs decreases in Instances DP4TP8 (with 8 A100 each) and DP32TP1 (with only 1 A100 each), the maximum supportable text length decreases, while performance on  text generation tasks of standard length improves. 

This phenomenon originates from the different computational characteristics of attention and non-attention layers across context lengths of high dynamic range. As shown in \autoref{fig:challenge1}(b), the tensor size of attention layers grows steadily with the context length, thus requiring more memory space and a higher degree of parallelism, typically necessitating deployment across more GPUs. For example, supporting a single text generation task of 1000K length would require at least 8 A100 GPUs. In contrast, the tensor size of non-attention layers does not change with text length; hence, no GPU number increase is needed. Traditional LLM parallel strategies~\cite{megatronlm,huang2019gpipe} do not differentiate between attention and non-attention layers, applying static model splits such as tensor or pipeline parallelism indiscriminately. This non-differentiation can lead to non-attention layers being mapped to an excessive number of GPUs, potentially reducing computational performance due to over-segmentation. As shown in \autoref{fig:challenge1}(c), for a text generation task of 1000K length, the performance of non-attention layers on an 8 GPU instances is only about one-third of that on a single GPU instance.


\para{Observation 2: When handling tasks with long context lengths, the computational utilization of GPU significantly decreases, and for short contexts, there is insufficient GPU memory utilization. This leads to instances typically exhibiting low computational or low memory utilization during the service process, as illustrated in \autoref{fig:challenge2}(a).} To illustrate this, we analyzed the performance of the decode phase for a single A100 LLaMA2-7B instance in a simplified scenario where all requests have the same context length. The results, depicted in \autoref{fig:challenge2}(b), show that when the context length is 20, the memory can support up to 1800 requests. However, this situation indicates over batching; compared to a batch size of 800, there is hardly any improvement in GPU compute utilization, suggesting that the additional 1000 requests do not contribute to performance improvement but instead significantly increase the request's latency. In the appropriate case with a batch size of 800, the actual GPU memory utilization is only 42\%. Given that single A100 GPUs have a fixed total memory capacity of 80GB, as the context length of requests increases, the system is forced to handle smaller batches. Specifically, when the context length reaches 20K, the maximum batch size is limited to 3, leading to GPU compute utilization being only one percent of what it is when the context length is 20.


This issue is rooted in the stark contrasts in computational characteristics and resource demands between attention and non-attention layers. Non-attention layers utilize weight parameters that can be shared across all input vectors for requests. This capability allows the system to transform `matrix-vector' multiplications (GEMV) into more efficient `matrix-matrix' multiplications (GEMM) as batch sizes increase, significantly improving the computation-to-memory ratio and achieving high computational utilization, as shown in \autoref{fig:challenge2}(c). With the continued growth in context length, LLM service systems are forced to reduce batch sizes to free up memory to accommodate the increasing KVCache of the attention layers, resulting in decreased GPU compute utilization. However, in previous systems~\cite{yu2022orca,kwon2023efficient,agrawal2023sarathi,sheng2023flexgen}, requests could only utilize the fixed resources available within their instance, limiting the system's ability to adapt to highly dynamic resource demands. 


In summary, these observations highlight the need for adaptive parallelism and resource management strategies to efficiently handle the varying demands of different context lengths, optimizing both computational and memory resource utilization across large-scale clusters.

\section{System Overview}

\begin{figure}[t!]
  \centering
  \includegraphics[width=1\linewidth]{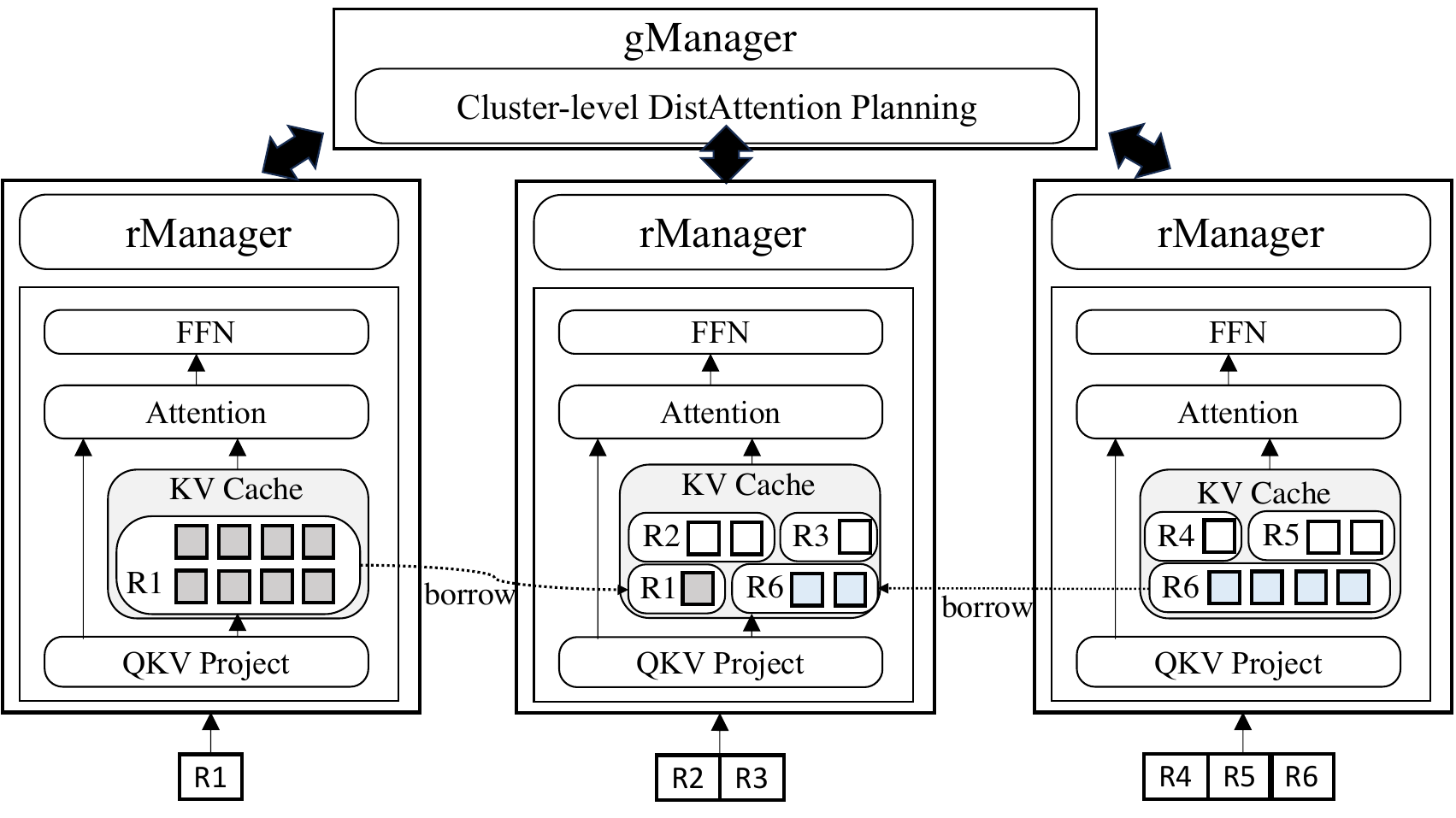}
  \caption{~\sysname~ System Overview}
  \label{fig:sys}
\end{figure}

The key concept of \sysname{} is to distribute the attention computation and KVCache beyond the boundaries of LLM inference instances, in order to leverage the resources of the entire GPU cluster, as illustrated in ~\autoref{fig:sys}. This idea disaggregates the attention layers from the non-attention layers, allowing them to employ independent parallel strategies and resource scheduling policies. It further enhances the scheduling strategy's ability to effectively manage the computation and memory of GPU resources at the cluster level.

\sysname{} accomplishes the design goal through three main system innovations. First, we introduce \attename{}, a novel attention mechanism that subdivides attention computations across GPUs, meanwhile avoiding KVCache transfer at decoding.
\attename{} is mathematically equivalent to the common attention modules, such as multi-head attention, multi-query attention, and  grouped-query attention\cite{shazeer2019fast, bert}. The new distributed attention mechanism allows partition the attention in arbitrary size of sequence length and introduces negligible output data to be transferred for the other layers. Therefore, attention can be efficiently disaggregated in a scalable way (Section~\ref{sec:algorithm}).
Second, \attename{} can be utilized in multiple  ways to resolve the resource contention and low efficiency of processing long requests and small requests thereby to significantly  improve the cluster throughput.  
We model the major attention and non-attention cost of using \attename{} to formulate the aggregated cluster throughput. Achieving the optimal cluster throughput is costly as LLM serving is dynamic and unpredictable. \sysname{} includes a greedy scheduling policy based on our empirical study, approaching the improved cluster throughput and efficiency (Section~\ref{sec:optimize}). 
Third, \sysname{} introduces a new centralized controller, gManager, to host the scheduling policy and coordinate the dynamic inter-instance KVCache tracking and migration. For scalability and fault tolerance, gManager works with a series of rManagers in a distributed architecture through a set of newly defined protocol (Section~\ref{sec:protocol}). 

\section{\attename{}~}
\label{sec:algorithm}

\begin{figure}[t!]
  \centering
  \includegraphics[width=0.9\linewidth]{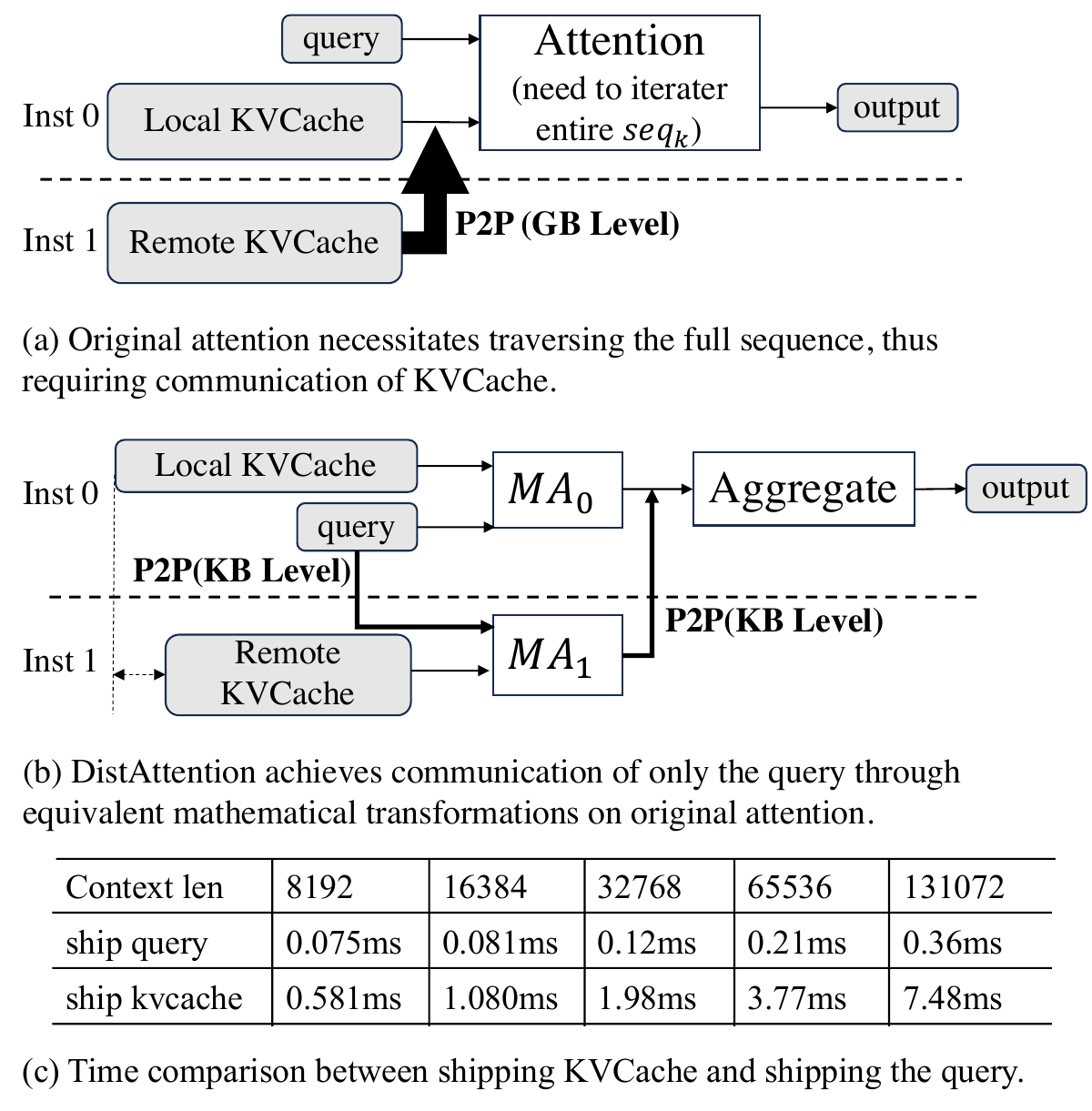}
  \caption{~\attename{} reduces  the communication overhead through equivalent mathematical transformations. 
  }
  \label{fig:dynblkatten}
\end{figure}

To achieve dynamic and flexible management of the KVCache, we propose ~\attename{}, a method that subdivides attention and the KVCache into regular small sub-blocks, thereby allowing the attention layer to be efficiently distributed and computed across multiple instances. Unlike traditional model parallelism methods, ~\attename{} is characterized by its slicing along the dynamic sequence dimension of the KVCache, allowing newly generated tokens in the auto-regressive process to be flexibly grouped, scheduled, and computed.
Although the KVCache tensor in the original attention can also be partitioned along the sequence dimension, the complex computation pattern of attention means that direct partition introduces significant communication overhead, greatly affecting the computational efficiency of distributed Attention. Inspired by online softmax\cite{rabe2021self}, \attename{} successfully addresses this issue through an equivalent mathematical transformation on the original attention.  ~\autoref{eq:attention} shows the original computation formula of attention,  requiring  calculating the maximum attention score ($m_g$ in ~\autoref{eq:attention}) across all sequences and summing the intermediate results along the sequence dimension, thus necessitating the entire KVCache of all sequences. If KVCache is directly partitioned and stored in a distributed manner, it would necessitate transferring the KVCache from remote instances back to the local machine for each attention computation. As illustrated in ~\autoref{fig:dynblkatten}(a), given the substantial size of the KVCache, each decoding step needs to transfer GBs or even TBs of data, significantly impacting the performance of distributed attention computation.
\begin{equation}
\small
\begin{aligned}
m_{g}=max(QK_{1},...,QK_{seq}) \notag 
\label{eq:max}
\end{aligned}
\end{equation}
\begin{equation}
\small
\begin{aligned}
    \text{Attention}(Q, K, V) = \sum_{i=1}^{seq} \frac{\exp({QK_{i}^T}-m_{g})}{\sum_{j=1}^{seq} \exp({QK_{j}^T}-m_{g})}V_{i} 
\label{eq:attention}
\end{aligned}
\end{equation}
\attename{}'s equivalent mathematical transformation on the original attention avoids the need to perform max and summation operations across all sequences. It allows each instance to execute the max and summation operations locally on partial KVCache data with partial sequence length $seq_p$. As shown in ~\autoref{eq:microattention}, MicroAttention ($MA$) refers to the partial attention computations that result from the partition and can be distributed across various instances for computation. Consequently, for each distributed computation of attention, instances  need to transfer the query vector along with only two float values, $e_j$ and $m_j$.
\begin{equation}
\small
\begin{aligned}
m_{j}=max(QK_{1},...,QK_{seq_{p}}), e_{j} = {\sum_{i=1}^{seq_{p}} \exp({QK_{i}^T}-m_{j})} \notag
\label{eq:max}
\end{aligned}
\end{equation}
\begin{equation}
\small
\begin{aligned}
    MA_{j}(Q, K, V) = \sum_{i=1}^{seq_{p}}( {\exp({QK_{i}^T}-m_{j})}V_{i}) 
\label{eq:microattention}
\end{aligned}
\end{equation}
\begin{equation}
\small
\begin{aligned}
    m_{g} = max(m_{1},...,m_{b}), e_{g}=\sum_{j=1}^{b}e_{j}\exp(m_{j}-m_{g})\notag
\label{eq:max_global}
\end{aligned}
\end{equation}
\begin{equation}
\small
\begin{aligned}
\text{Attention}(Q, K, V)=\sum_{j=1}^{b}\frac{MA_{j}\exp(m_{j}-m_{g})}{e_{g}}
\label{eq:max_global}
\end{aligned}
\end{equation}
The intermediate results computed by the $b$ remote instances are then transferred back to the local instance for aggregation (\autoref{eq:max_global}) to arrive at an outcome equivalent to the original attention, as indicated in ~\autoref{fig:dynblkatten}(b). Because the computation FLOPs for aggregation is less than 1\% of the total MA computational load, this overhead is virtually negligible. Since the query round trip involves only a few KBs of data, ~\attename{} substantially reduces the data communication overhead, as depicted in ~\autoref{fig:dynblkatten}(c).

\section{Cluster-scale Throughput Optimization}
\label{sec:optimize}

\subsection{Overview}

\attename{} allows \oursys{} to place and compute a single request across multiple instances. This is not merely a means to enable serving extremely long requests beyond a single instance's capacity; we show that, from a cluster perspective, \attename{} is a powerful weapon as it greatly enlarges the request scheduling space across instances and can improve cluster-wide throughput.
In particular, with \attename{}, \oursys{} is no longer limited to scheduling KVCache of each whole request on an instance; instead, \oursys{} can schedule any arbitrary sub-blocks of request's KVCache onto instances, which represents a much finer scheduling granularity and higher flexibility than existing systems. 

\begin{figure}[t]
  \centering
  \includegraphics[width=0.95\linewidth]{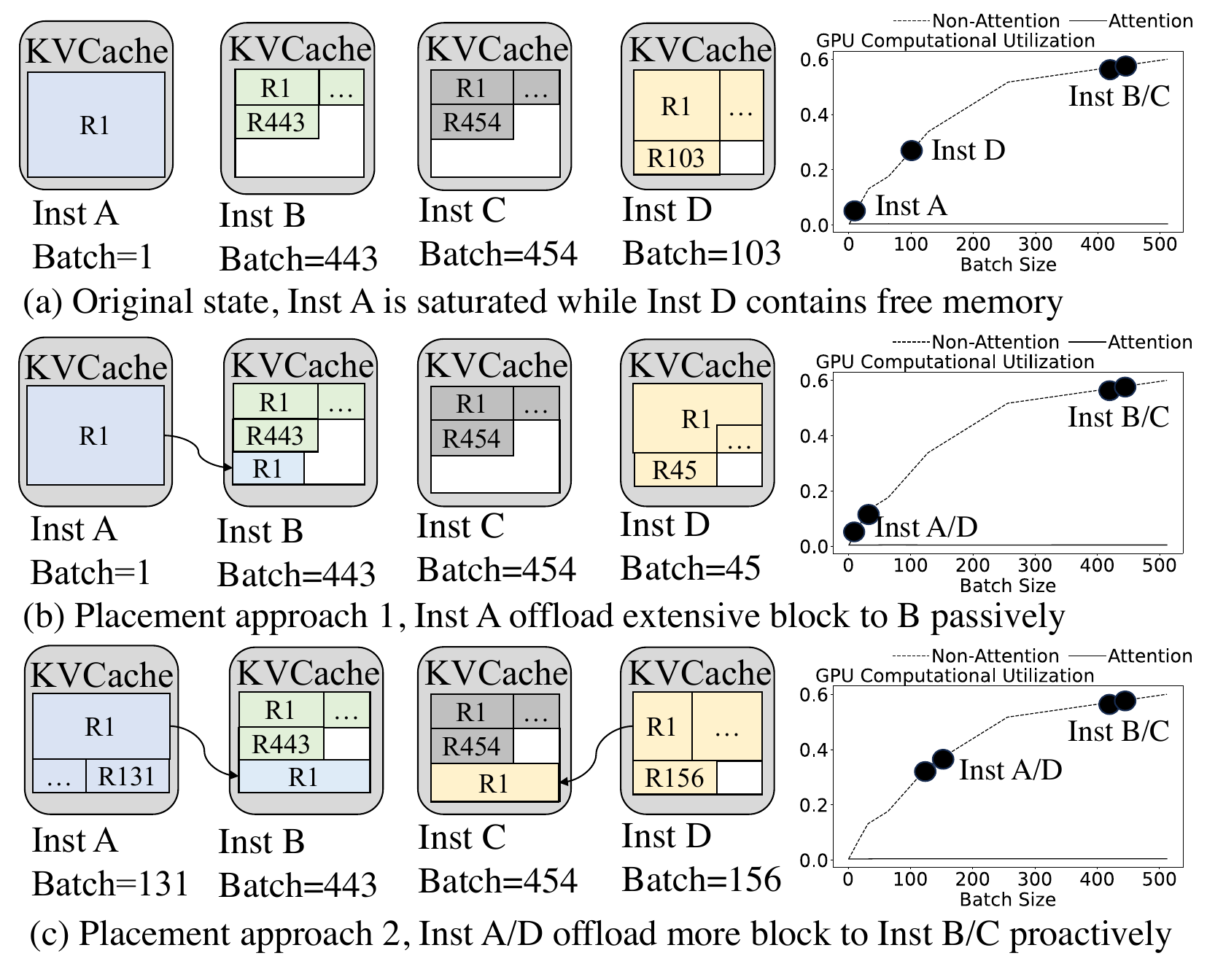}
  \caption{Effects of two different placement methods}
  \label{fig:overview-opt}
\end{figure}

Such sub-block level scheduling is beneficial because \oursys{} can better balance the KVCache and the batch sizes across instances, thereby maximizing memory and compute utilization simultaneously. Specifically, \oursys{} maintains balanced batch sizes by controlling the number of sub-blocks on each instance, preventing individual requests from occupying too much memory and decreasing the batch size. \autoref{fig:overview-opt} shows an intuitive example.
\autoref{fig:overview-opt}(a) demonstrates the initial status on the KVCache distribution of four serving instances. Instance A is processing a long request that saturates all memory space, and Instance D is processing a long request however with some available GPU memory left. Instances B and C are handling short requests, and despite high batch sizes, they still have ample memory space. \autoref{fig:overview-opt}(b) shows a strawman placement strategy that
only places the newly generated blocks onto the instance with the most remaining space when the length of a long request exceeds the memory capacity of an instance. While this approach can support long requests that exceed the resource capacity of a single instance, the cluster as a whole maintains a relatively low throughput: the batch size for Instance A remains at 1, and the newly generated attention sub-blocks of the long request in Instance D compete with local short requests, reducing the batch size and resulting in low GPU compute utilization. The second placement method (\autoref{fig:overview-opt}(c)) proactively places more sub-blocks onto other instances with sufficient available space before the length of a long request exceeds the instance memory capacity. As shown in ~\autoref{fig:overview-opt}(c), Instances A and D proactively place more attention sub-blocks to Instances B and C, freeing up memory space to handle more short requests, thereby increasing the batch size. Compared to the first method, this proactive placement balances the batch sizes among instances, improving the overall cluster throughput. 

The insights summarized from the above study inspire us to devise a reasonable scheduling method for placing attention sub-blocks to enhance the overall throughput of the cluster as much as possible. Our scheduling method must address three main issues: (1) If an instance offloads part of its KVCache to remote, how to determine a proper size? What is the performance gain and overhead? (2) If an instance lends some space out, how much space should be used? 
(3) Given that there are numerous instances, 
how to decide the borrow-lend relationship that maximizes the overall performance? 
\subsection{Debtors and Creditors}

To address these problems, we discuss the performance impacts between instances that borrow or lend memory spaces. We refer to instances that borrow memory from others as \textit{debtors}, such as Instances A and D, and instances that lend memory as \textit{creditors}, such as Instances B and C in ~\autoref{fig:overview-opt}. We do not allow an instance to act both as a debtor and a creditor simultaneously. Whenever a debtor has free local space (e.g., a request retires and frees up its memory), it prefers to increase its batch size, or to retrieve attention sub-blocks that were previously offloaded instead of lending it to others. Similarly, a creditor, lacking sufficient local space, will reclaim the memory that it has previously lent out.

\begin{figure}[t!]
  \centering
  \includegraphics[width=0.95\linewidth]{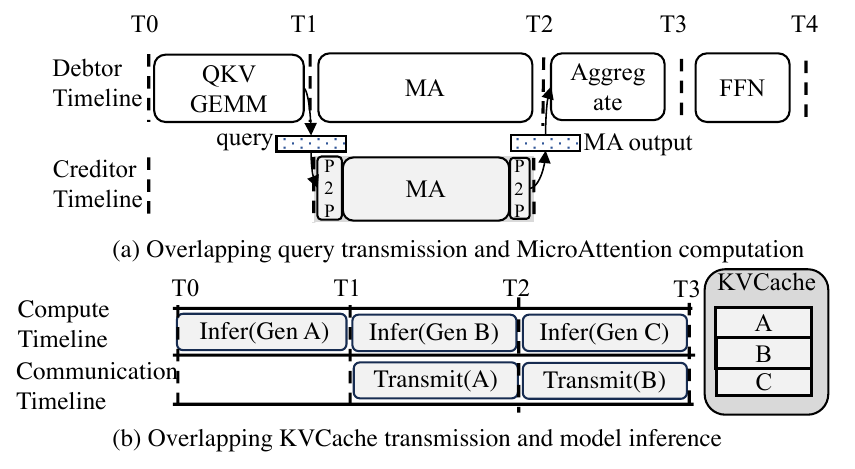}
  \caption{Communication Overlapping Optimization}
  \label{fig:overlap_optimize}
\end{figure}

\begin{figure*}[t]
  \centering
  \includegraphics[width=1\linewidth]{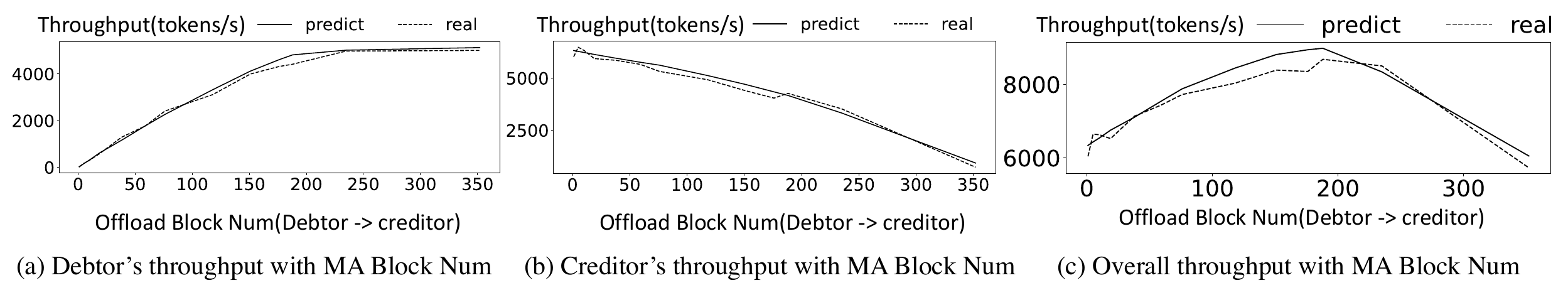}
  \caption{Experiment is conducted using LLAMA2-7B model on two A100 instances. One instance acts as a creditor executing normal-length requests (500 tokens on averag). The other acts as a debtor, processing a long context (1000K tokens). }
  \label{fig:cluster_throughput}
\end{figure*}

\subsubsection{Debtors}

Debtors borrow memory space from one or more creditors to store portions of their KVCache. This operation has both positive and negative impacts on their performance. The primary benefit is that offloading part of the attention computation to creditors reduces the time debtors spend generating tokens, and the freed-up space allows higher batch size, thereby improving generation throughput.

This approach also introduces several challenges. First, debtors must collect the partial attention computation results from creditors to complete the attention calculation.  If creditors compute an excessive number of MicroAttention (MA) operations, it may lead to idle waiting for the debtors and thus reduce performance. As shown in \autoref{fig:overlap_optimize}(a), to mitigate this issue, our scheduling policy aims to limit the size of KVCache on remote instances so that the remote computation and transmission can be entirely covered by local computations. Secondly, transferring the KVCache to creditors is time-consuming. To minimize the impact of KVCache transfer on LLM inference performance, as shown in \autoref{fig:overlap_optimize}(b), we overlap the transfer with local model inference. 

\para{Micro-benchmarking and performance analysis.} Figure~\ref{fig:cluster_throughput}(a) depicts the debtor's throughput as a function of the number of KVCache blocks moved to the creditor, represented by the dashed line. As more KVCache are offloaded, the debtor's throughput greatly increases because the debtor enjoys a boost in performance when its batch size is very low. As the batched requests approach the computational limits of the system, this trend eventually plateaus. 

\subsubsection{Creditors}
Creditors lend their excess memory space to one or multiple debtors. Due to the additional computation of partial attention for debtors, the performance of creditors’ local request services is negatively affected.

\para{Micro-benchmarking and performance analysis.} \autoref{fig:cluster_throughput}(b) reveals a slow but steady performance degradation with more KVCache moved to the creditor. When the transferred KVCache exceeds the surplus memory space of the creditor, the batch size of the creditor is reduced, resulting in an even steeper decline in performance.

\subsubsection{Overall cluster throughput} 
The overall system throughput is obtained by summing the throughput of all instances in the cluster. Our optimization goal is to find an effective KVCache placement strategy that enhances the throughput of the entire system. For example, \autoref{fig:cluster_throughput}(c) presents the aggregated performance of a micro-benchmarking test for a debtor and a creditor. As the KVCache is transferred, the total throughput increases from approximately 6500 tokens/second to about 8800 tokens/second. As more MA blocks are moved to the creditor, the overall throughput sharply declines. Under this configuration, the system achieves maximum aggregate throughput when 200 blocks of KVCache are transferred.



\para{Complexity analysis.} However, determining a KVCache placement schedule for the real cluster is very difficult because the design space is prohibitively complex. Considering a cluster with $N$ instances as debtors and $M$ ones as creditors, we refer the number of surplus memory blocks in each creditor as $Y_i$. For each block, there are $N+1$ possible options: offering it to one of the $N$ debtors or to not lend it out at all. Supposing all memory blocks make the decision independently, the search space can be $(N+1)^{\sum_{i=1}^{M}Y_i}$. However, the blocks within each creditor are homogeneous. After deduplication, the final search space size is as follows:
\begin{equation}
\begin{aligned}
\frac{(N+1)^{\sum_{i=1}^{M}Y_i}}{\prod_{i=1}^{M}Y_{i}!}
\label{eq:search space}
\end{aligned}
\end{equation}

Such a huge design space makes it impractical to figure out the optimal cluster throughput during runtime. Next, to avoid the overhead of empirical measurements, we have constructed a performance model to predict the overall cluster throughput for a given placement strategy, and we propose an optimization algorithm based on this model.

\subsection{Scheduling Algorithm}
\label{sec:schedule-algo}

To efficiently figure out an efficient KVCache placement schedule during runtime, we propose a greedy algorithm based on a performance model.

\para{Performance modeling.} ~\autoref{eq:transformer} outlines a general analytical model for single transformer layer, comprising both non-attention and attention layers. The computational load of all non-attention layers, denoted as $W(\beta)$,  is primarily influenced by batch size $\beta$. The GPU's real performance(FLOPs/s), denoted as $f(\beta)$, is closely tied to batch size and can be experimentally measured. The workload of attention layers is dictated by the requests' length $S$. Since attention layers cannot benefit from batching, their GPU performance, denoted as $g(S)$, typically remains constant and is also ascertainable through experimental methods.

\begin{equation}
\small
\begin{aligned}
T^{lyr}(\beta, S)=& T^{natn}(\beta)+T^{atn}(S) 
=& \frac{W(\beta)}{f(\beta)}+\sum_{r=1}^{\beta}\frac{S^r}{g(S)}
\label{eq:transformer}
\end{aligned}
\end{equation}

\autoref{eq:inst_0_offload_latency} extends this performance model to specifically address the roles of debtors and creditors within the system. In \oursys{}, a debtor can offload KVCaches of size $K^d$ to creditors, allowing an increase in its batch size to $\beta'$. Meanwhile, a creditor may allocate space for KVCaches of size $K^c$ to debtors while maintaining its original batch size $\beta$. 


\begin{equation}
\small
\begin{aligned}
T_{dbt}^{lyr}(\beta', K^d)=& T^{lyr}-\frac{K^d} {g(S)},
T_{cdt}^{lyr}(\beta, K^c)=& T^{lyr}+\frac{K^c} {g(S)}
\label{eq:inst_0_offload_latency}
\end{aligned}
\end{equation}

Combining the formulations above, per-instance throughput (a.k.a, tokens per second) is as $TPS = \frac{\beta}{n\cdot T^{layer}}$, where $n$ represents the number of transformer layers of an LLM. For a cluster deployed with $M$ instances, the overall aggregated throughput equals to the sum of the $TPS$ of all instances: 
\begin{equation}
\small
\begin{aligned}
TPS_{cluster} = \sum_{i=1}^{M}TPS_i
\label{eq:cluster_tps}
\end{aligned}
\end{equation}
We have validated the accuracy of the performance model, which is shown in ~\autoref{fig:cluster_throughput}. The results predicted by the performance model are consistent with the real measurements. Having obtained the formula for calculating the overall cluster throughput, we can estimate the cluster throughput with any specific schedules.

\para{Greedy Algorithm.} 
We propose a greedy algorithm to maximize the cluster throughput with \attename{}~ approximately. Our algorithm is founded on a principle: pairing overloaded debtor instances with the free creditor to continuously perform load balancing scheduling, thereby enhancing overall throughput.

As illustrated in Algorithm~\autoref{algo:schedule}, we select instances with batch sizes smaller than the empirical threshold $\beta^{thres}$ as debtor instances, while those with memory utilization rates below the empirical threshold $U^{thres}$ serve as creditor instances. 
As small batch size instances are with great performance potential empirically, debtors are processed in ascending order according to their batch sizes.
At each round, the longest request $r$ is selected from the debtor, as well as the creditor with maximal available memory.
The possible block number to move for request $r$ is explored to estimate the potential throughput gain. 
Specifically, we first establish the upper limit on the number of MA blocks that can be offloaded, which corresponds to the number of MA blocks for request $r$. 
Under the constraint of maximum number of MA Blocks, $Block_{max}$, the performance model of aggregated throughput of two instances (as illustrated in ~\autoref{eq:cluster_tps}) is utilized to determine the number of offload blocks between 0 and $Block_{max}$. 
For each debtor, the algorithm loops the creditors in orders until no performance gain can be achieved from the memory block movement.
In \oursys{}, the algorithm acts periodically and retrospecively to adapt to the dynamic and unpredictable serving load of LLM.

\begin{algorithm}[t!]
\caption{Cluster-level DistAttention Scheduling}
\label{algo:schedule}
\small
\begin{algorithmic}[1]
\State Collect debtors with small batch size $D^i \in \{I^i, \beta^i \leq \beta^{thres} \}$
\State Sort debtors in increased batch size order $\langle I^i, \beta^i\rangle$
\State Collect creditors with low memory util $C^i \in \{I^i, U^i \leq U^{thres} \}$
\State Sort creditors in increased memory util order $\langle I^i, U^i\rangle$
\For {$D^i \in D$}
    \State $r$=pick\_longest\_request($D^i$)
    \State $Block_{max}$ = get\_block\_num($r$)
    \For {$C^j \in C$}
        \For{$k \in$ range(0,$Block_{max}$)}
            \State $\langle perf,k \rangle$ = perf\_model\_throughput\_esti($k$, $D^i$, $C^j$) 
        \EndFor
        \State $Block_{best}$ = pick\_max\_perf($\langle perf,k \rangle$)
        \If{$Block_{best} <= 0$}
            \State break;
        \EndIf
        \State move\_kvcache($D^i$, $C^j$, $Block_{best}$)
        \State update\_and\_sort\_mem\_util($C$)
        \State $Block_{max}-=Block_{best}$
    \EndFor
\EndFor
\end{algorithmic}
\end{algorithm}

\section{System Design}
\label{sec:protocol}

\subsection{gManager and rManager}

To realize the global planning described in \autoref{sec:optimize}, \sysname{} employs a centralized manager termed the \textit{gManager} to maintain a global view of instance status and make the request and KVCache placement decisions. The gManager tracks the KVCache placement on the instances of each request, maintained in the \emph{request placement map}, where each entry represents (part of) the KVCache memory usage of a request on a certain instance. A request is allowed to be distributed on multiple instances, i.e., having multiple entries. One of the entries/instances of a request is marked as the debtor instance of it.
Utilizing this data, gManager tracks the current status of request placements and then derives a new expected placement status and the transition plan.

Considering the rapid changing memory usages of requests,
tracking the status of every single request precisely in the gManager would be prohibitively expensive.
\sysname{} develops a distributed and coordinated system architecture to implement the global planning efficiently.
As shown in Figure \ref{fig:sys}, \sysname{} 
introduces a series of distributed \textit{rManager}s co-located with the instances.
The gManager and rManagers work in a \emph{loosely-coordinated} manner. 
That is, instead of keeping the global view in sync with the real request statuses, it relies on periodic heartbeat signals from the rManager of each instance to convey updates about the KVCache memory usages of requests on it. This approach reduces the overhead of the global planning and enhances system performance. After receiving a full update from all instances in each round, the gManager calculates a status transition plan and finally instructs the rManagers to move the KVCaches if needed.

Under this architecture, \sysname{} needs to deal with the potential staleness of the global view during the periodic updates. In normal cases, the running batch on each instance and its KVCache memory usage keep growing as the decoding computation proceeds after a periodic status update.
Moreover, due to the continuous batching behavior, the memory usage may experience a steeper change when certain requests complete or new requests join.
We design a protocol among the gManager and rManagers to implement this interaction.
\subsection{Protocol}

\begin{figure}[t!]
  \centering
  \includegraphics[width=0.91\linewidth]{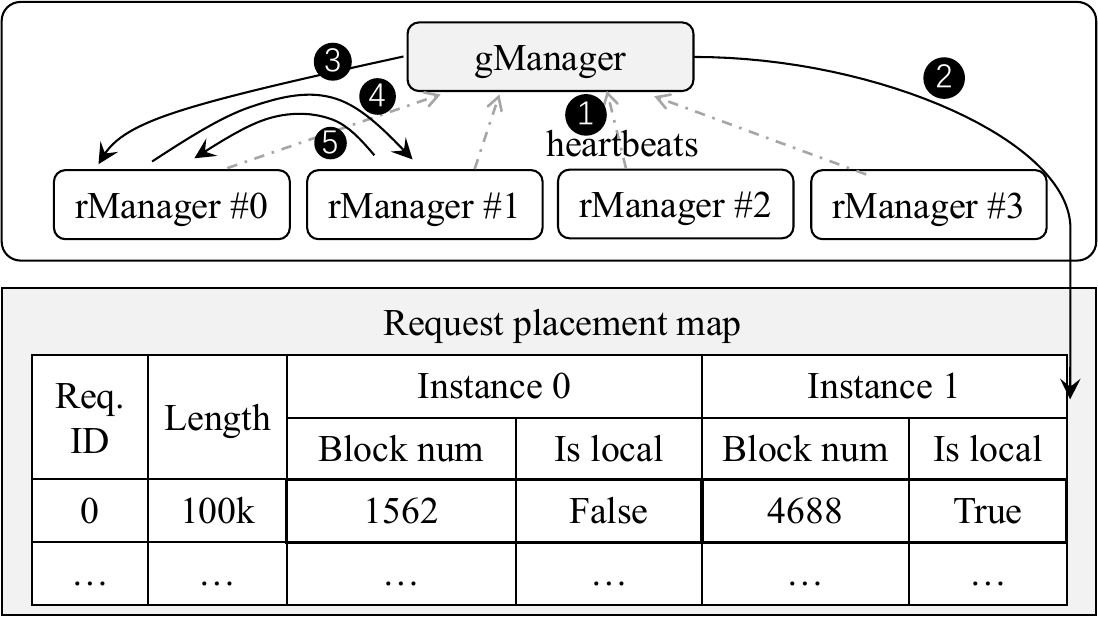}
  \caption{ Overall workflow of ~\sysname’s protocol}
  \label{fig:protocol}
\end{figure}

\autoref{fig:protocol} shows the overall workflow of \sysname{}'s protocol.
We summarize the APIs used in the protocol in Listing 1.
Each rManager reports its local status using the \texttt{heartbeat} API~\circled{1}, which includes an array of the request placement entries (shown by the \texttt{RequestPlacementEntry} struct). Note that when a request locates in multiple instances, it is possible that its status on certain instances do not change in a period if the newly generated KVCaches are not placed on them. Therefore, in normal cases, the rManager only sends the entries that have changed since the last update to the gManager. An exception is that when initializing a new gManager (e.g., after a failover), the rManager will send the full information to help the gManager construct the initial status.
The gManager updates these entries into the global request placement map accordingly~\circled{2}.
The gManager then dispatches its request placement decisions using the \texttt{move\_kvcache} API,
which instructs an instance to move a certain amount of KVCache blocks to a destination instance~\circled{3}.

\begin{lstlisting}[language=python, caption=Infinite-LLM APIs]
class RequestPlacementEntry:
    req_id:int, inst_id:int, num_blocks:int, local:bool
heartbeat(List[RequestPlacementEntry]) -> None
move_kvcache(req_id:int, num_blocks:int, dst_inst:int) -> None
try_move_kvcache(req_id:int, num_blocks:int)->bool
\end{lstlisting}

Considering the potential staleness of the request placement map, the instruction from the gManager to move KVCache to another instance could be infeasible --- for example, when the KVCache on the destination instance grows and the memory space becomes insufficient.
Therefore, \sysname{} further provides the \texttt{try\_move\_kvcache} API for the source instance to try to reserve space on the destination instance before transferring the real KVCache data~\circled{4}. On the destination side, it may receive multiple concurrent \texttt{try\_move\_kvcache} calls from other instances. The destination instance uses a \emph{first-come-first-serve} policy to decide the allocation among these competing candidates; if the total space is not enough for satisfying all of the them, the destination instance will reject some of them. The desination instance responds with a boolean value to identify whether this KVCache movement is allowed or rejected~\circled{5}. If it is allowed, the source instance proceeds to transfer the data; otherwise, it simply waits for further instructions from the gManager in future rounds, which will have captured the latest cluster status.

\section{Evaluation}
\label{sec:eval}

\subsection{Experimental Setup}


\para{Environment.}
We deploy \sysname~ on a cluster with 4 nodes and 32 GPUs. Each node has 8xNVIDIA A100 (80GB) GPUs. The GPUs are connected via NVLink (600GB/s) within each node and via Ethernet (125MB/s) across nodes.

\para{Models.}   
Since most LLM models have similar backbone Transformer block, we choose one representative model family, LLaMA2\cite{llama2} for evaluation. The LLaMA2 family contains three different model sizes: 7B, 13B and 70B. They use two popular attention architectures; the 7B and 13B models utilize Multi-head Attention (MHA)\cite{bert}, while the 70B model employs Grouped-Query Attention (GQA)\cite{shazeer2019fast}.

\para{Traces.} We generate 9 traces with different context length ranges and length distributions to comprehensively evaluate \sysname{}'s end-to-end performance. Traces 0-2, marked as ``S'' (short) in \autoref{tab:dataset-scenarios}, have relatively short sequence lengths that are guaranteed to fit in each instance when using vLLM (i.e., vLLM-multi). 
Requests of trace 0 come from the open-source dataset ShareGPT4\cite{sharegpt4}, which contains conversations of GPT4 service. To assess the impact of different length distributions, particularly the variance of sequence lengths, we select a subset of data from ShareGPT4 to construct traces 1 and 2 with reduced standard deviations. Traces 3-8, marked as ``L'' (long) in \autoref{tab:dataset-scenarios}, are used to evaluate \sysname{} using larger context length ranges, where requests of trace 3 come from open-source dataset L-Eval\cite{an2023leval}, and traces 4-8 are from the distribution of long requests from our online service.
In each experiment, we assign an arrival time to each request using a Poisson distribution using varying request rates.

\begin{table}[t]
\centering
\footnotesize
\begin{tabular}{ccccc}
\toprule
\multicolumn{2}{c}{Trace} & Range   & Avg.   & SD        \\
\midrule
\multirow{3}{*}{S}   & 0  & 1-60k   & 1233   & 7785.68         \\
                     & 1  & 1-60k   & 712    & 5531.4         \\
                     & 2  & 1-60k   & 469    & 3506.36         \\ \midrule
\multirow{6}{*}{L}   & 3  & 1-200k  & 56362  & 28787.78  \\
                     & 4  & 1-280k  & 75650  & 39479.42  \\
                     & 5  & 1-600k  & 160239 & 87906.67  \\
                     & 6  & 1-480k  & 128804 & 70647.93  \\
                     & 7  & 1-1200k & 293945 & 172169.14 \\
                     & 8  & 1-2000k & 498609 & 261817.24 \\
                     \bottomrule
\end{tabular}
\caption{Ranges, average values, and standard deviations (SDs) of context lengths of the traces.}\label{tab:dataset-scenarios}
\end{table}

\para{Comparison.} Our evaluation focuses on comparing \sysname{} with static model parallelism and resource planning. To this end, we use vLLM\cite{kwon2023efficient}, a state-of-the-art LLM serving engine using static model parallelism, as the primary baseline. Specifically, we compare the following approaches:

\begin{itemize}
    \item \sysname{}: Given the total cluster resources, \sysname{} divides them into multiple model instances using an appropriate parallelism configuration (for non-attention computation) while scaling the attention computation across the instances. \sysname{} dispatches each request to the instance with the most free GPU memory.
    \item \emph{vLLM-multi} (vLLM-M): vLLM with the same number and parallelism configuration of the instances as \sysname{}. It might fail to run some long requests due to limited per-instance memory capacity. vLLM-M uses the same dispatching policy as \sysname{}.
    \item \emph{vLLM-single} (vLLM-S): a single instance containing all the cluster resources, so that vLLM can support the same sequence length ranges as \sysname{}. Note that vLLM only supports tensor parallelism, which is known to be less efficient than pipeline parallelism when distributed across machines\cite{megatronlm}. We implement pipeline parallelism in vLLM for cross-machine communication when this instance needs to be distributed on multiple machines.
\end{itemize}


\subsection{Context Length Performance}

We first benchmark the performance of \sysname{} and the baselines when running requests with different context lengths. We use six context length ranges and three models. For each model and range, we test three specific context lengths: (1) a short context length (1k); (2) a length slightly exceeding the maximum length that vLLM-multi can support; and (3) the maximum length that \sysname{} supports given the cluster resources. For each data point, we measure the throughput of a largest batch of requests given the context length.
\begin{figure*}[t!] 
    \centering
    \includegraphics[width=0.99\linewidth]{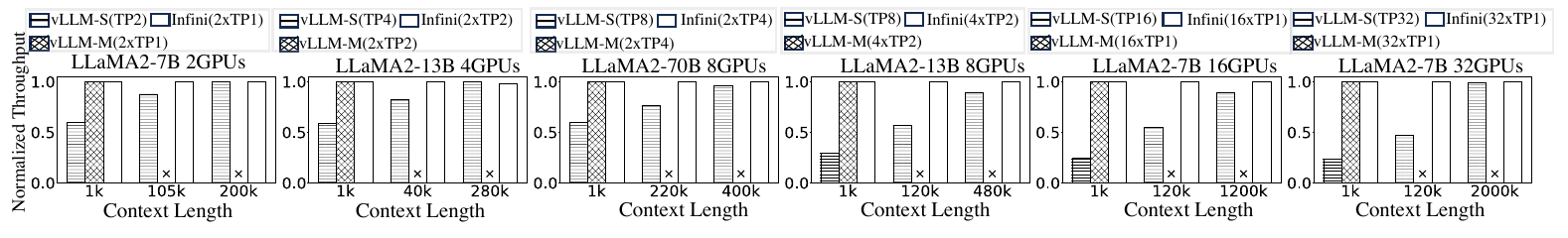} 
    \caption{Context Length Performance}
    \label{fig:context_length_benchmark}
\end{figure*}

As shown in \autoref{fig:context_length_benchmark},
\sysname{} achieves the best of both long and short sequences. Compared to vLLM-multi, which is limited by per-instance memory, \sysname{} supports substantially longer context (2x-19x) while achieving comparable throughput on short sequence lengths. This improvement is attributed to \sysname's ability to efficiently coordinate memory and computation usage across all instances, while vLLM-multi is limited to the instance's private resource. Compared to vLLM-single, \sysname~ obtains 1.4x-5.3x higher throughput on short lengths while sustaining similar longest context lengths. This is because \sysname~ can maintain an efficient model parallelism strategy for FNN computations while vLLM-single has to partition the model into smaller segments across more GPUs, which results in lower GPU computation utilization for the Non-Attention parts and more communication overhead.

\subsection{End-to-end Serving Performance}

\begin{figure}[t!] 
    \centering
    \begin{subfigure}{\columnwidth}
    \includegraphics[width=1\linewidth]{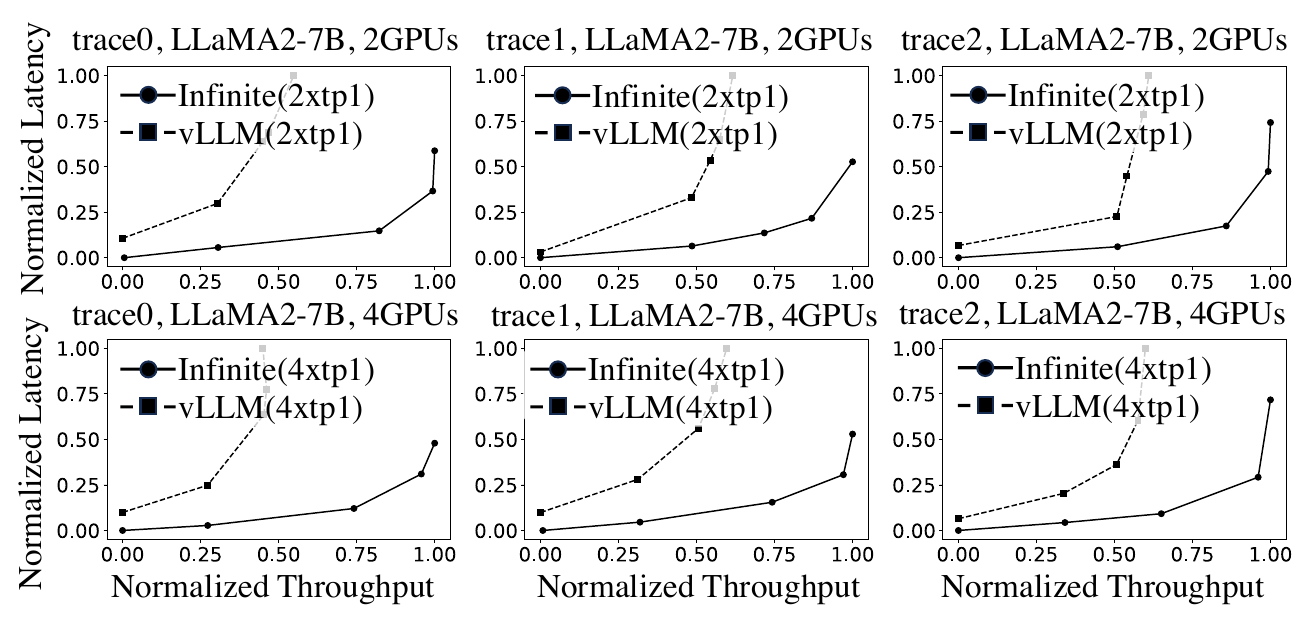}
    \caption{Comparison with vLLM-M}\label{fig:end_to_end_serving_small}
    \end{subfigure}
    
    \begin{subfigure}{\columnwidth}
    \includegraphics[width=1\linewidth]{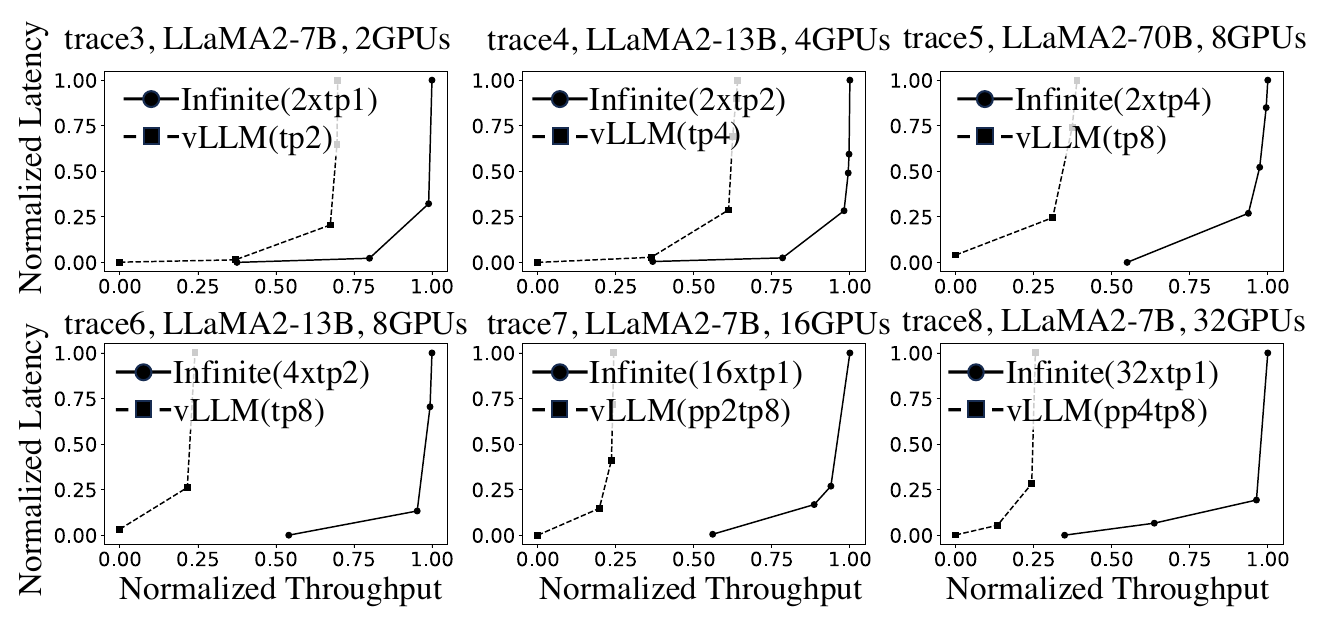}
    \caption{Comparison with vLLM-S}\label{fig:end_to_end_serving_large}
    \end{subfigure}
    \caption{End-to-end serving performance.}
\end{figure}

\para{Comparison with multiple small instances.}
We first compare \sysname{} with vLLM-multi, which launches multiple instances with the same parallelism configuration as \sysname{}.
We conduct six experiments using Traces 0-2, where the sequence lengths won't exceed the limit of each single instance of vLLM.
\autoref{fig:end_to_end_serving_small} shows the throughput-latency variation when using different request rates. We compare the maximum achieved throughputs of \sysname{} and vLLM.
The results demonstrate that ~\sysname~ gets a throughput improvement of approximately 1.35x-1.73x over vLLM.
We further examine how the number of instances and traces' context length distribution affect performance gains. As depicted in the six sub-figures of ~\autoref{fig:end_to_end_serving_small}, from left to right, the standard deviation of the traces' context length distribution decreases while the number of instances increases from top to bottom. We observe that the performance gains rise with the standard deviation (indicating a more uneven length distribution) and the number of instances . This is because a more uneven length distribution or a larger number of instances lead to greater variance in resource demands among different instances, enhancing the benefits of unified resource management across all instances.

\para{Comparison with a single large instance.}
We use Traces 3-8 with longer context lengths to compare \sysname{} with vLLM-S, which allocates all GPUs to a single instance to accommodate sufficiently long sequences.
The results shown in ~\autoref{fig:end_to_end_serving_large} indicate that ~\sysname~ gets a 1.4x to 3.4x throughput gain over vLLM. From top to bottom and left to right in ~\autoref{fig:end_to_end_serving_large}, we observe that ~\sysname's performance gains grow with the context length range expanding. This is attributed to vLLM's static model parallelism fragmenting the model across more GPUs, leading to reduced efficiency in the non-attention segments and significantly lowering the system’s capability to process shorter request efficiently, whereas ~\sysname~ maintains an appropriate model parallelism strategy for the non-attention part, thereby preserving their performance.



\subsection{Micro-benchmarks}

\begin{figure}[t!] 
    \centering
    \includegraphics[width=0.9\linewidth]{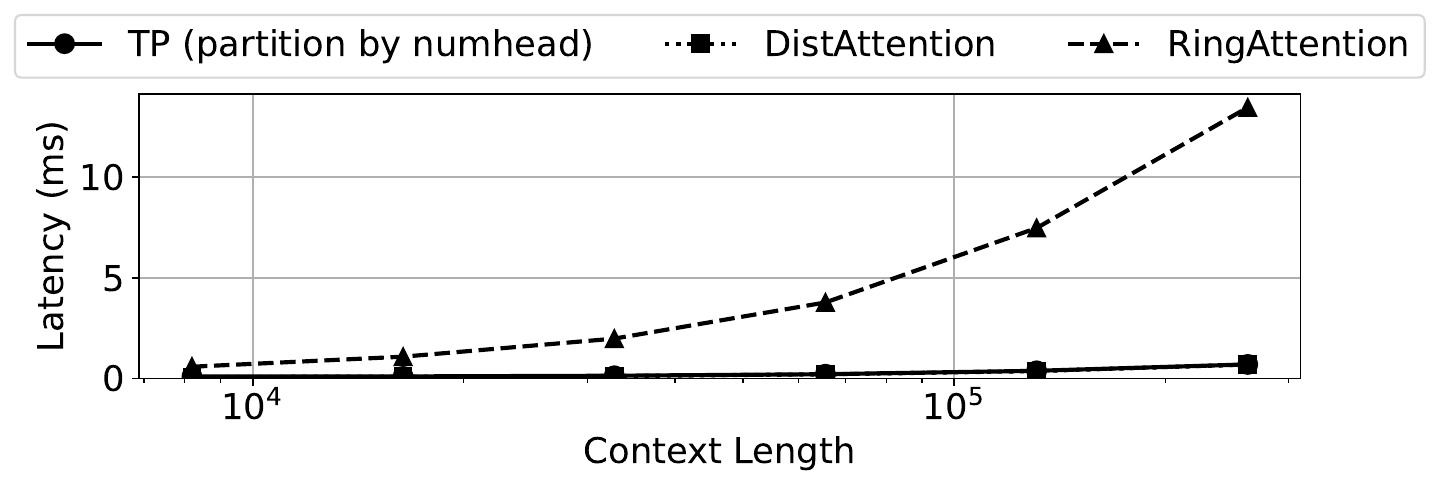} 
    \caption{Comparison of distributed attention methods.}
    \label{fig:local_remote_atten_overlap}
\end{figure}

\para{Comparison with other long-context attention methods.}
We compared the performance of \attename{}, RingAttention, and TP (partition by numer of heads) within the context range from 4K to 256K, where the Attention computation is based on the dimensions of LLaMA2-13B with four GPUs. As shown in ~\autoref{fig:local_remote_atten_overlap}, the results show that \attename{} is 1\%-25\% faster than TP due to its lower communication overhead. Compared to RingAttention, \attename{} is 7.7x-19.8x faster owing to the significantly higher communication overhead of RingAttention, which involves the transfer of large KVCache (MB to GB), whereas \attename{} transmits very small-sized queries (KB).



\begin{figure}[t!] 
    \centering
    \includegraphics[width=0.9\linewidth]{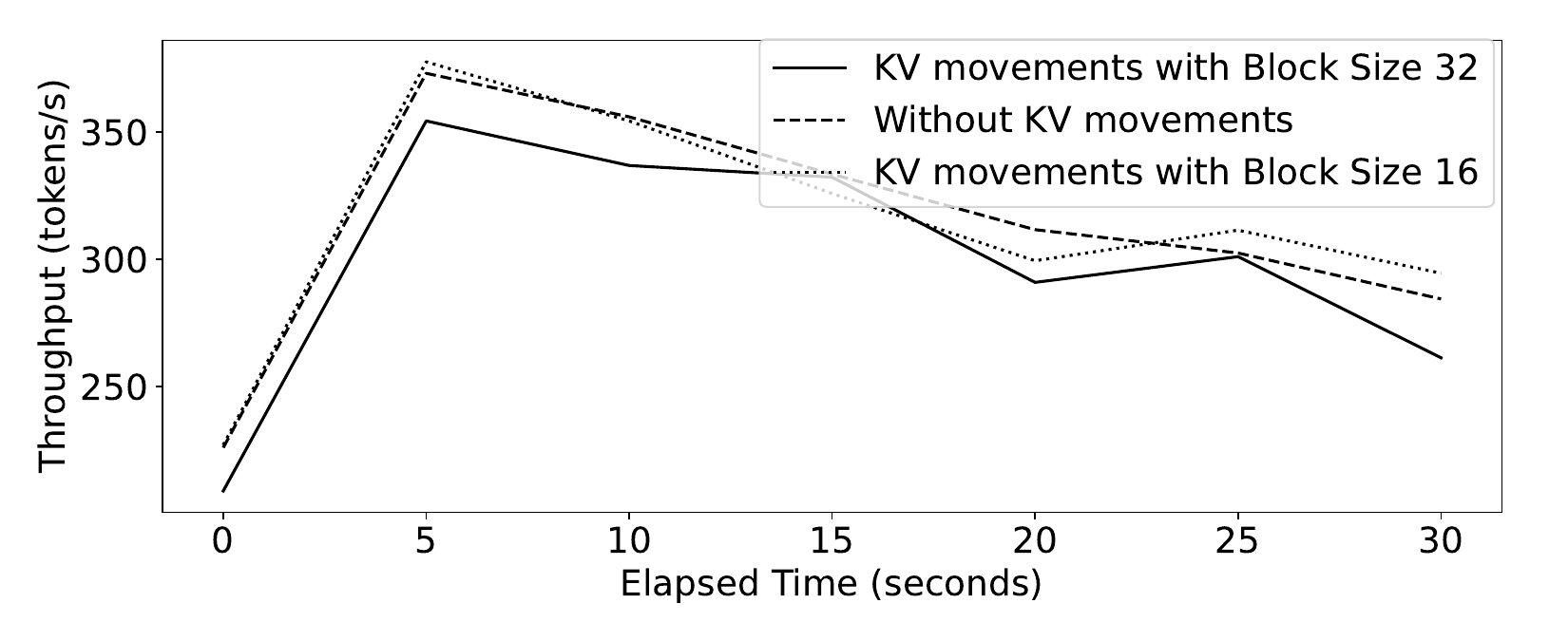}
    \caption{Overhead of KVCache movement.}
\label{fig:migration_comm_overhead}
\end{figure}

\para{Overhead of KVCache movement.}
We improve cluster-scale throughput by scheduling ~\attename{} across all instances. To reduce the overhead of KVCache movement between instances, ~\sysname~ overlaps the movement with model computation. To evaluate the impact of movement communication on instance throughput, we compared the instance throughput with movement enabled to that with movement disabled. It's important to note that movement in this experiment does not change the batch size, hence any fluctuations in the throughput curve are due to the communication costs of movement. Results shown in ~\autoref{fig:migration_comm_overhead}(a) indicate that instance throughput decreased by 8.6\% when moving 32 tokens per decode step. When moving 16 tokens per decode step, the throughput of instance was identical to the instance with movement turned off. Therefore, when the movement size is set to 16, communication can overlap well with computation without affecting the instance's performance.

\section{Related Work}
\label{sec:related}

\para{LLM inference system.}
ORCA~\cite{yu2022orca} introduced iteration-level scheduling which greatly enhances the computation and memory utilization in batching inference. vLLM~\cite{kwon2023efficient} further proposed PagedAttention to address the memory wastage due to fragmentation. DeepSpeed-FastGen\cite{fastgen2023} proposed a novel prompt and generation composition strategy called Dynamic SplitFuse (or Sarathi\cite{agrawal2023sarathi}) to further enhance system throughput.   DistServe\cite{zhong2024distserve} proposed to disaggregate the prefill and decode stages to different instances to avoid their interference.
Despite these novel systems solve many problems and achieve outstanding results, the dynamic problem along with the need to support exceptionally long context lengths still remains an unresolved challenge.

\para{Long-context LLM.}Works like FlashAttention\cite{dao2022flashattention} and FlashDecoding\cite{fdc} focus on optimizing the performance of attention in long sequences. They enhance the compute-to-memory ratio and SM (Streaming Multiprocessors) parallelism of Attention on a single GPU by addressing data dependency issues. However, they do not take into account the communication overhead in multi-GPU settings and cannot be directly applied to scenarios involving multiple GPUs. To train LLM with long context, some research work~\cite{liu2023ring, brandon2023striped, li2023lightseq, li2021sequence} has introduced the method of context parallelism to partition the computation in sequence dimension. Ring Attention~\cite{liu2023ring,liu2023blockwise} distributes long sequences across multiple devices, with the intent of fully overlapping the communication of KV blocks with the computation of blockwise attention. 
Those methods are designed for training, which is a poor fit to the highly dynamic characteristic in LLM inference decoding phase, causing substantial overhead to transfer KVCache across devices at each iteration.
Another thread to address the challenge of oversized KVCache for long-context inference is to utilize sparse KVCaches such as Sliding Window Attention~\cite{mistral,sparsetransformers,beltagy2020longformer}, H2O~\cite{h20} and StreamingLLM~\cite{xiao2023efficient}, which compromises with the potential accuracy loss, because of the KVCache eviction. 
\oursys{} supports long-context LLM serving by introducing a new scalable distributed attention mechanism, \attename{}. Attention can be disaggregated from the model inference, therefore to schedule across multiple serving instances, both for computation and KVCache management. \attename{} retains equivalence to the original attention thereby be harmless to model accuracy.

\para{Scheduling.}
To improve throughput and latency of LLM serving, serveral systems~\cite{sun2024llumnix,sheng2024fairness, lin2024parrot, fu2024serverlessllm} have been proposed to optimize for request scheduling across multiple model instances.
Llumnix\cite{sun2024llumnix} dynamically reschedules requests across multiple instances at runtime to deal with the heterogeneity and unpredictability of requests. Parrot\cite{lin2024parrot} uncovers the dependencies and commonalities among LLM requests, thus creating a new space for enhancing the end-to-end performance of LLM applications. However, the previous work is limited to scheduling
each whole request on an instance, facing issue of low GPU utilization due to dynamic request length.
\oursys{} can schedule any arbitrary sub-sequences of requests onto instances, representing a much finer scheduling granularity and higher flexibility than existing systems.

\section{Conclusion}
\label{sec:conclusion}
In this paper, we have 
presented \oursys{}, a novel LLM service system designed for
managing highly dynamic context lengths in LLM requests. 
Through \oursys{}, we have revealed the highly dynamic characteristic within LLM requests and advocated attention disaggregation to be a common technology for LLM serving.
In particular, we have introduced a novel system architecture both efficient and scalable for all LLM requests, proposed a scheduling policy to saturate the computation and bandwidth of GPU simultaneously, and shown significant improvement through extensive evaluations on representative real traces. Going forward, we hope that \oursys{} can become a common foundation toward AGI for both the research community and industry, inspiring future advancements in LLM serving.

\bibliographystyle{plain}
\bibliography{ref}

\end{document}